\documentclass[twocolumn,prl,aps,floatfix,superscriptaddress,amssymb]{revtex4}
\usepackage{epsfig}

\begin{document}

\title{Edge stability, reconstruction, zero-energy states and magnetism in triangular graphene quantum dots with zigzag edges}

\author{O.~Voznyy}
\affiliation{Institute for Microstructural Sciences, National Research Council of Canada
, Ottawa, Canada}

\author{A.~D.~G\"u\c{c}l\"u}
\affiliation{Institute for Microstructural Sciences, National Research Council of Canada
, Ottawa, Canada}

\author{P.~Potasz}
\affiliation{Institute for Microstructural Sciences, National Research Council of Canada
, Ottawa, Canada}
\affiliation{Institute of Physics, Wroclaw University of Technology, Wroclaw, Poland}

\author{P.~Hawrylak}
\affiliation{Institute for Microstructural Sciences, National Research Council of Canada
, Ottawa, Canada}

\date{\today}

\begin{abstract}
We present the results of ab-initio density functional theory based calculations of the stability and
reconstruction of zigzag edges in triangular graphene quantum dots.
We show that, while the reconstructed pentagon-heptagon zigzag edge structure is more
stable in the absence of hydrogen, ideal zigzag edges are energetically
favored by hydrogen passivation. Zero-energy band exists in both structures
when passivated by hydrogen, however in case of pentagon-heptagon zigzag, this
band is found to have stronger dispersion,  leading to the loss of net
magnetization. 
\end{abstract}

\maketitle

Graphene, a single layer honeycomb lattice of carbon atoms, exhibits
fascinating properties due to relativistic-like nature of quasiparticle
dispersion close to the Fermi level\cite{NGM+04,NGM+05,ZTS+05,ZGG+06,NGP+09}.
Graphene's potential for nanoelectronics applications motivated considerable
amount of research in graphene nanoribbons\cite{NFD+96,WFA+99,WaSS+08,YCL08,CRJ+10}
and, more recently, graphene quantum
dots\cite{WSG08,WRA+09,AB08,LSB09,ZCP08,YNW06,Eza07,Eza08,FP07,WMK08,AHM08,GPV+09,gucluarxiv,PGH10}. In
such low-dimensional structures, the character of the edge drastically affects
the electronic properties  near the Fermi
level\cite{YNW06,KFE+06,YK08,RL09}. In particular, assuming stable zigzag edge
in  triangular graphene quantum dots (TGQDs), tight-binding and density functional theory based methods
predicted collapse of the energy spectrum near the Fermi level
to a shell of degenerate states, isolated from the rest of the
spectrum by a well defined
gap\cite{Eza07,Eza08,FP07,WMK08,AHM08,GPV+09,PGH10,YNW06}. It was shown that
in this band of degenerate states, strong electron-electron 
interactions lead to ferromagnetism with peculiar magnetic\cite{GPV+09,WMK08} and
optical\cite{gucluarxiv} properties.
Recently, the potential of nanoscale graphene flakes for use in photovoltaics was demonstrated\cite{flakesPV_NL2010}.
Demonstration of multiexciton generation in carbon nanotubes\cite{GaborMEG,RabaniMEG} 
suggests its possibility in other graphene-related materials. 
Combined with the possibility to control the bandgap 
and with the presence of intermediate band in the gap, it makes TGQDs an attractive material 
for third generation solar cells\cite{Nozik_review2010}.

The stability\cite{WaSS+08,KMH08,GME+09,KMH09,EFJ+10}, 
control over\cite{JHM+09,CMS+09,CRJ+10}, and
physical effects\cite{YNW06,KFE+06,YK08,RL09} of edges  in graphene
structures were studied experimentally and theoretically. It was predicted\cite{WaSS+08,KMH08,KMH09}
that in nanoribbons the zigzag configuration ($ZZ$)
is not necessarly the most stable, but a transition to reconstructed edge,
terminated by pentagon-heptagon pairs ($ZZ_{57}$), can occur. The $ZZ_{57}$
reconstruction  was also observed experimentally\cite{GME+09,KMH09, CMA+09} in
graphene boundaries. In general, confining Dirac fermions in two-dimensions
requires the understanding of the role of the edges in finite 
area graphene based nanostructures.

In this work, using ab-initio methods we investigate the reconstruction of the edges 
and corners of TGQDs 
and their effect on zero-energy states and magnetism. 
We focus on the competition between $ZZ$ and $ZZ_{57}$ configurations which
were shown to be the most stable in previous
works\cite{WaSS+08,KMH08,KMH09}. For hydrogen passivated edges, we find $ZZ$ structure 
to be the most stable. We show that, in TGQDs, the reconstruction can occur in
various ways due to the presence of corners, and the most stable reconstructions
cause breaking of the reflection symmetry of the triangular
structure. Moreover, in hydrogen passivated $ZZ_{57}$ structures, zero energy
states survive, although the ferromagnetism is destroyed due to stronger
dispersion. 

\begin{figure}
\epsfig{file=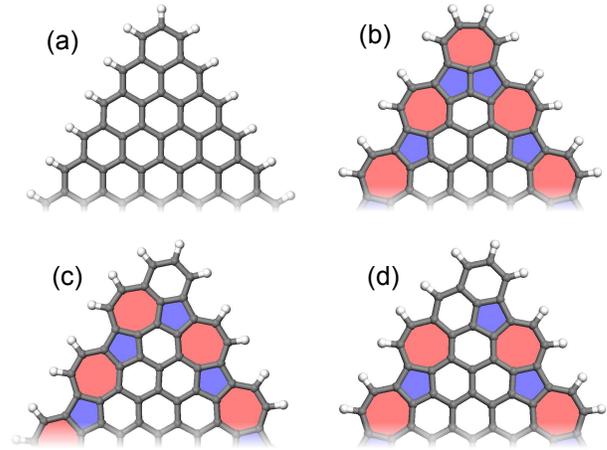,width=3.2in}
\caption{\label{fig:zztypes} (Color online) Triangular graphene quantum dot edge
configurations considered in this work: (a) Ideal zigzag edges, $ZZ$, 
(b) $ZZ_{57}$ reconstruction with pentagon-heptagon-pentagon corner
configuration, (c) $ZZ_{57}$ reconstruction with heptagon-hexagon-pentagon corner, 
and (d) $ZZ_{57}$ reconstruction with hexagon-hexagon-pentagon corner.
}
\end{figure}

Calculations have been performed within the density functional theory (DFT)
approach as implemented in the SIESTA code\cite{siesta}.
We have used the generalized gradient approximation (GGA) with the
Perdew-Burke-Ernzerhof  exchange-correlation functional (PBE)\cite{pbe}, 
double-$\zeta$  plus polarization orbital (DZP) bases for all atoms 
(i.e. $2s$, $2p$ and $2d$ orbitals for carbon, thus, both $\sigma$- and $\pi$-bonds are included on equal footing) 
and Troullier-Martins norm-conserving pseudopotentials to represent the cores, 
 300 Ry real space mesh cutoff for charge density,  and a supercell with at
least 20 {\AA} of vacuum between the periodic images of the TGQDs.  
Geometries were optimized until the forces on atoms below 40 meV/{\AA} were reached  and
exactly the same geometries were used for the comparison of total energies of
the  ferromagnetic (FM) and antiferromagnetic (AFM) configurations. 
Our optimized C-C bond length for bulk graphene of 1.424{\AA} 
overestimates the experimental value by $\sim$3\%, typical for GGA.


Figure 1 shows different TGQD structures considered in this work.  
Depending on the parity of the amount
of atoms in the edge of TGQD,  the requirement of the $ZZ_{57}$ reconstruction
of the edge leads to several possible structures  of the triangle corner,
presented in Fig.1(b)-(d) (for the sake of comparison of total energies  we
investigate only those reconstructions conserving the amount of atoms).  The
three rings at the corner can have 5-7-5 (Fig.1(b)),
7-6-5  (Fig.1(c)), or 6-6-5 arrangements (Fig.1(d)). 
Among reconstructed corners,  only the structure in
Fig.1(b) conserves the mirror symmetry of the TGQD,  however, according to our
calculations, it is the least stable  due to strong distortion of the corner
cells.  Thus, in the remainder of the work we will be presenting results
utilizing configuration shown in Fig.1(c) for even, and that in Fig.1(d) for
odd number of atoms $n$ on a side of the triangle.

\begin{figure}
\epsfig{file=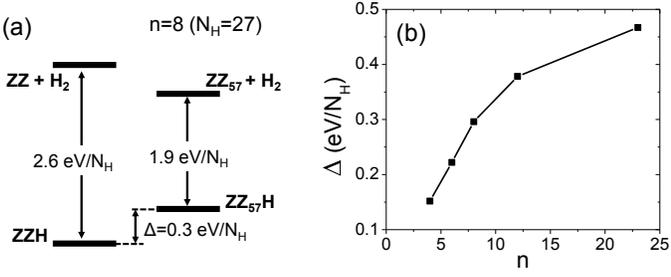,width=3.5in}
\caption{\label{fig:EtotvsNH} 
  (a) Relative total energies of hydrogen-passivated and non-passivated TGQDs 
  with reconstructed and non-reconstructed edges. Presented values are per hydrogen atom 
  for  the case of $n=8$ ($N_H=27$)
  (b) Total energy difference between hydrogen-passivated
  $ZZ_{57}$ and $ZZ$ configurations as a function of number of atoms on a side of the triangle.
}
\end{figure}
Passivation by hydrogen is an important
requirement for the observation of attractive properties of TGQDs
(zero-energy band of nonbonding states and magnetization of edges) which arise  
from topological frustration of the $\pi$-bonds \cite{WMK08}. 
Our calculations show that without hydrogen passivation, the $\pi$-bonds hybridize with the $\sigma$-bonds 
on the edge, destroying the zero-energy band.
In Fig.2(a) we address the stability of hydrogen passivation for $ZZ$ and $ZZ_{57}$ edges 
on the example of  a triangle with $n=8$ atoms on a side (97 carbon atoms total, 
number of passivating hydrogens $N_H=3n+3=27$).  
For hydrogen-passivated structures, $ZZH$ is $0.3$ eV per hydrogen atom more stable than $ZZ_{57}H$ 
since in the latter structure the angles between the $\sigma$-bonds significantly deviate from 
the ideal 120$^\circ$ and the total energy is affected by strain. 
In the absence of hydrogen, however, the structure has to passivate the dangling $\sigma$-bonds
by itself, e.g. by reconstructing the edge. Indeed, the $ZZ_{57}$ reconstruction becomes $0.4$ eV more stable.
It is important to note that hydrogen passivation is a favorable
process for both structures even relative to the formation of $H_2$ molecules
and not only atomic hydrogen (Fig.2(a)). Same conclusions hold for larger TGQDs as well. 
These results are also consistent with the ones for graphene nanoribbons \cite{WaSS+08,KMH08}.  
Thus, we will present further only the results for hydrogen-passivated structures and will omit the index $H$ 
(i.e. use $ZZ$ instead of $ZZH$).
In Fig.2(b), we investigate the relative stability of  hydrogen-passivated 
$ZZ$ and $ZZ_{57}$ structures as a function of linear size of the triangles. The largest TGQD that
we have studied has $n=23$ atoms on a side of the triangle (622 carbon atoms total). 
The fact that energy per edge atom 
increases with size signifies that the bulk limit has not been reached yet. 
Clearly the $ZZ$  structure remains the ground state for the range of sizes  studied here.

\begin{figure}
\epsfig{file=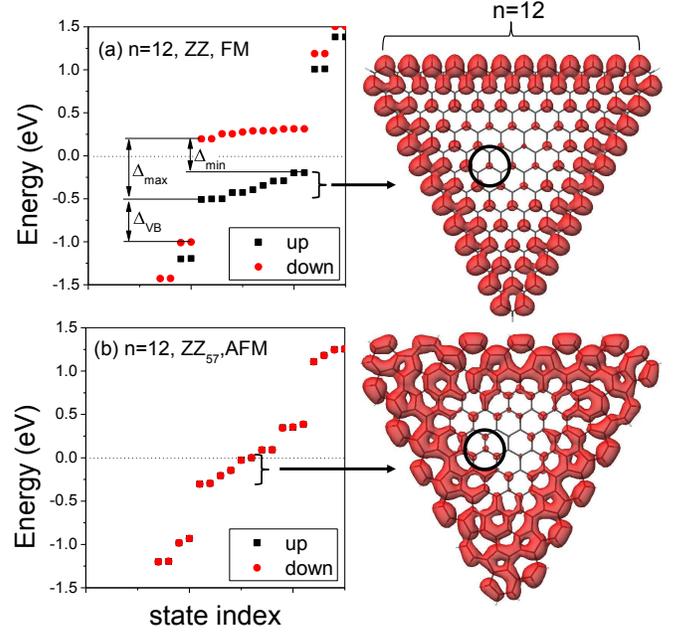,width=3.5in}
\caption{\label{fig:spectra} (Color online) Energy spectra of the ground states for (a) $ZZ$
  and (b) $ZZ_{57}$ configurations for a hydrogen-passivated
  triangular dot with $n=12$. Spin-up states are shown in black squares and spin-down
  states are shown in red circles. On the right hand side, charge
  densities of the filled part of zero-energy bands are shown.
  Circular outlines show the population of only one sublattice in $ZZ$ structure and both sublattices in $ZZ_{57}$.
  }
\end{figure}

It was shown previously that the number of states in the zero-energy band equals 
the difference between the number of atoms in A and B graphene sublattices. 
Zero-energy  states are localized exclusively on the sublattice to which the $ZZ$ edges belong 
and are exactly degenerate within the nearest neighbor tight-binding model\cite{WMK08,FP07,PGH10}.
Fig.3 compares the DFT electronic spectra near the Fermi level for the ground states of 
hydrogen-passivated unreconstructed ($ZZ$) and reconstructed ($ZZ_{57}$) TGQDs with $n=12$.
Introduction of the $ZZ_{57}$ edge reconstruction smears the distinction between sublattices.
Nevertheless, the zero-energy band survives in a reconstructed ($ZZ_{57}$) TGQD.
As can be seen from the electronic density of occupied states of the shell, 
they are still predominantly localized on the edges. 
Moreover, the number of zero-energy states remains the same. However, the dispersion of this band
increases almost three-fold due to reduction of the structure symmetry. 
Lifting of the band degeneracy becomes observed 
even in the nearest neighbor tight-binding model with equal hoppings, and is more pronounced for the structures 
in Fig.1(c) and (d) which additionally lift the reflection symmetry present in Fig.1(b).

\begin{figure}
\epsfig{file=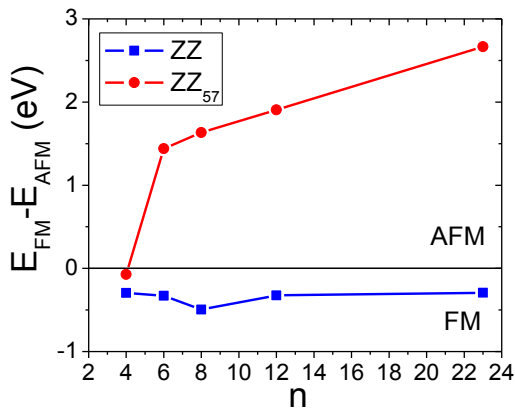,width=2.8in}
\caption{\label{fig:FMvsAFM} (Color online) Total energy difference between
ferromagnetic and antiferromagnetic states as a function of the size of the
triangle for hydrogen-passivated $ZZ$ (blue squares) and $ZZ_{57}$ (red circles). For $ZZ$ the
ground state is ferromagnetic for
all sizes studied, while for $ZZ_{57}$ it is antiferromagnetic for $n>4$
}
\end{figure}

Magnetization of the $ZZ$ configuration was investigated in detail
through meanfield\cite{FP07,WMK08} and exact diagonalization\cite{GPV+09}
calculations. It was shown that the electrons in the zero-energy band are spin-polarized. 
The up- and down-spin edge states are split around the Fermi level such that only 
up-spin states are filled. Our calculated dispersion of the up-spin states is $\sim$0.03
eV/state (Fig.3(a)). On the other hand, the ground state of $ZZ_{57}$ configuration is
antiferromagnetic, i.e. there is no splitting between the up- and down-spin states (Fig.3(b)).
Nevertheless, calculations for the ferromagnetic $ZZ_{57}$ can still be performed
by adjusting the Fermi level for up and down spins independently.
The energy sectrum obtained in such a way is similar to the one for $ZZ$ but with negative $\Delta_{min}$.
The interplay of the $\Delta_{max}$ and the spin-up band dispersion in such FM calculation 
can be monitored to predict whether the ground state will be 
ferromagnetic ($\Delta_{min}>0$) or antiferromagnetic ($\Delta_{min}<0$). 
Apart from the significant increase of the band dispersion,
we note that the spin up-down splitting $\Delta_{max}$ reduces by a factor of two in $ZZ_{57}$ structure. 
One can see from charge density plot in Fig.3(b) that zero-energy states can now populate both A and B sublattices 
even close to the center of the dot (see outlined regions).
We speculate that the resulting reduction in the the peak charge density on each site 
is responsible for the reduced on-site repulsion between spin-up and spin-down electrons.
Stronger dispersion and reduced up-down spin splitting favor kinetic energy minimization 
versus exchange energy and destroy the ferromagnetism in $ZZ_{57}$.
It should be noted that partial polarization can still be possible in $ZZ_{57}$. 
Particularly, we observed it for structures with symmetric corners (Fig.1(b)) which exhibit smaller dispersion.

Our conclusions based on the analysis of the energy spectra 
are supported by the total energy calculations depicted in Fig.4 .
For $ZZ$ structure the gap $\Delta_{min}$ is always positive and the total energy of the 
FM configuration is lower than that of AFM (blue squares).
For $ZZ_{57}$ configuration, on the opposite, the ground state clearly remains AFM  for all sizes with the exception of the case with n=4. 
Here the band consists of only 3 states
and their dispersion cannot overcome the splitting between spin-up and spin-down states, resulting in
FM configuration beeing more stable.
The total energy difference between the FM and AFM configurations for $ZZ$, 
remains almost constant (in the range 0.3-0.5 eV) for the triangle sizes studied here, 
and reduces with size if divided by the number of edge atoms. 
Such a small value, comparable to the numerical accuracy of the method,
makes it difficult to make reliable predictions regarding magnetization of larger dots.

\begin{figure}
\epsfig{file=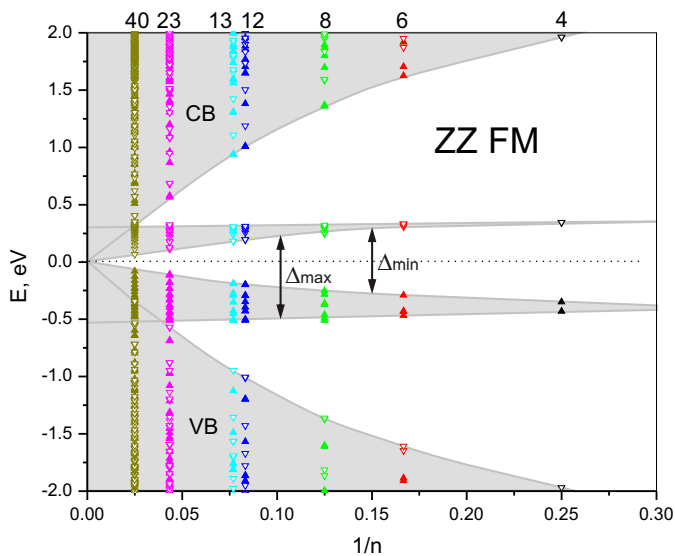,width=3.5in}
\caption{\label{fig:gapscaling} (Color online) Scaling of the energy gaps with the inverse linear size of $ZZ$ TGQDs. 
Full energy spectra of the structures calculated in present work are shown. 
Open symbols correspond to spin-down and filled ones to spin-up states.
}
\end{figure}
In order to investigate whether the magnetization of the edges would be preserved on mesoscale
we plot in Fig.5 the evolution of the energy spectra with the TGQD size. 
For this plot we performed an additional calculation for the case of $n=40$ (1761 carbon atoms total). 
We did not perform the geometry optimization for this case due to high computational cost,
however, based on the results for smaller structures we expect that this would have minor effect on the spectrum.
This alows us to notice the reduction of the splitting $\Delta_{max}$ 
between the spin-up and spin-down states with the growing size, not appreciated previously\cite{WMK08}.
Our GGA gap between zero-energy bands ($\Delta_{min}$)
and that between the valence and conduction bands
are larger than LDA gaps reported previously\cite{WMK08}, 
as also observed for graphene nanoribbons\cite{Rudberg_NanoLett07}.
Both gaps show sublinear behavior, complicating the extrapolation to triangles of infinite size.
This behavior, however, should change to linear for larger structures where the effect of edges reduces\cite{gucluarxiv},
converging both gaps to zero, as expected for Dirac fermions.
An important difference from the nearest neighbor tight-binding calculation\cite{PGH10} 
is the growing dispersion of the zero-energy bands.
Combined with the reducing valence-conduction gap, 
this leads to the overlap of the zero-energy band with the valence band even for finite sizes, 
as indeed observed for the $n=40$ case (see Fig.5), while in $ZZ_{57}$ structures it becomes visible already at $n=23$.
Nevertheless, it does not affect the magnetization of the edges, as indeed confirmed by our calculation for $n=40$,
and can be compared to a magnetization of the infinitely long hydrogen-passivated nanoribbons
where the edge state overlaps in energy with the valence band but in $k$-space those bands do not actually cross\cite{WaSS+08}.
Our results thus suggest that magnetization of the edges for infinitely large triangles 
survives in the limit of zero temperature.

In conclusion, we have investigated the edge stability, reconstructions, hydrogen passivation in triangular graphene quantum dots 
and their effect on electronic and magnetic properties of the dots as a function
of size. For hydrogen passivated edges, zigzag configuration is found to be
more stable than pentagon-heptagon edges. The reconstruction into
pentagon-heptagon zigzag edges involves a reconstruction of the triangle's
corners as well, and causes the reflection symmetry to be broken. Finite
magnetism is found to be destroyed in TGQDs with reconstructed ($ZZ_{57}$) edges, 
however, the band of zero-energy states survives.

{\it Acknowledgement}. The authors thank M. Korkusinski for fruitful discussions,
and NRC-CNRS CRP, Canadian Institute for
Advanced Research, Institute for Microstructural Sciences, and QuantumWorks
for support.

\vspace*{-0.22in}



\end{document}